# Field-normalized citation impact indicators and the choice of an appropriate counting method

Ludo Waltman and Nees Jan van Eck

Centre for Science and Technology Studies, Leiden University, The Netherlands
{waltmanlr, ecknjpvan}@cwts.leidenuniv.nl

Bibliometric studies often rely on field-normalized citation impact indicators in order to make comparisons between scientific fields. We discuss the connection between field normalization and the choice of a counting method for handling publications with multiple co-authors. Our focus is on the choice between full counting and fractional counting. Based on an extensive theoretical and empirical analysis, we argue that properly field-normalized results cannot be obtained when full counting is used. Fractional counting does provide results that are properly field normalized. We therefore recommend the use of fractional counting in bibliometric studies that require field normalization, especially in studies at the level of countries and research organizations. We also compare different variants of fractional counting. In general, it seems best to use either the author-level or the address-level variant of fractional counting.

## 1. Introduction

In discussions on bibliometric indicators, two topics that receive a considerable amount of attention are field normalization and counting methods. Field normalization is about the problem of correcting for differences in citation practices between scientific fields. The challenge is to develop citation-based indicators that allow for valid between-field comparisons. Counting methods are about the way in which co-authored publications are handled. For instance, if a publication is co-authored by two countries, should the publication be counted as a full publication for each country or should it be counted as half a publication for each country?

The topics of field normalization and counting methods are usually discussed separately from each other. However, we argue that there is a close connection between the two topics. Our argument is that proper field normalization is possible only if a suitable counting method is used. In particular, we claim that properly field-normalized results cannot be obtained when one uses the popular full counting method, in which co-authored publications are fully assigned to each co-author. The fractional counting method, which assigns co-authored publications fractionally to each co-author, does provide properly field-normalized results. The problem of full counting basically is that co-authored publications are counted multiple times, once for each co-author, which creates a bias in favor of fields in which there is a lot of co-authorship and in which co-authorship correlates with additional citations. This is the essence of the argument that we present in this paper. Our argument builds on an earlier paper (Waltman et al., 2012), but in the present paper we elaborate the argument in more detail and we also present an extensive empirical analysis.

The rest of this paper is organized as follows. In Section 2, we present the counting methods that we study in the paper. We discuss the connection between counting methods and field normalization in Section 3. We also introduce the concept of the full counting bonus in this section. This concept plays a key role in our ideas on counting methods. An empirical analysis of the full counting bonus is reported in Section 4. Empirical comparisons between different counting methods are presented in Section 5. In Section 6, we discuss some commonly used arguments in favor of full counting, and we provide a response to each of these arguments. Finally, we draw conclusions in Section 7.

## 2. Counting methods

In this section, we first provide an overview of the different counting methods that we consider in this paper. We then present a simple example in which the different counting methods are illustrated. This is followed by a discussion of the choice between a number of fractional counting variants. Finally, we review earlier work on counting methods.

### 2.1. Overview of counting methods

Our main focus in this paper is on the comparison between full counting and fractional counting. In the case of full counting, a publication is fully assigned to each co-author. For instance, a publication co-authored by four countries counts as a full publication for each of the four countries. In the fractional counting case, a publication is fractionally assigned to each co-author. The weight with which a publication is assigned to a co-author indicates the share of the publication allocated to that co-author. The sum of the weights of all co-authors of a publication equals one. An example of fractional counting is the situation in which a publication co-authored by four countries is assigned to each country with a weight of 1 / 4 = 0.25.

Fractional counting can be implemented in a number of different ways. In this paper, we distinguish between the following variants of fractional counting:

- *Author-level fractional counting*. Each author of a publication has equal weight.
- *Address-level fractional counting*. Each address listed in the address list of a publication has equal weight.
- *Organization-level fractional counting*. Each organization listed in the address list of a publication has equal weight.
- *Country-level fractional counting*. Each country listed in the address list of a publication has equal weight.

In addition to full and fractional counting, we also consider first author counting and corresponding author counting in some of the analyses presented in this paper. First author counting assigns a publication with a weight of one to the first author and with a weight of zero to each of the other authors. The underlying idea is that the first author of a publication often represents the most important contributor.[1] Corresponding author counting is similar to first author counting, but it assigns a publication with a weight of one to the corresponding author rather than to the first author. The other authors again have a weight of zero.

**2.2 Example**

To illustrate the different counting methods, we provide a simple example. We consider a publication that has five authors. The address list of the publication contains five addresses. Table 1 indicates which addresses belong to which authors. Table 2 shows the organization and the country mentioned in each of the addresses. Other address details, such as the department within the organization, the postal code, and the city, are not important in our example, and therefore these details are not provided in Table 2. We note that some authors in Table 1 have more than one address. These are authors with multiple affiliations. Based on Table 2, we observe that four different organizations and three different countries are listed in the address list of the publication that we consider. We also observe that address 1 and address 2 correspond with the same organization and with the same country. Although this is not visible in Table 2, other details of these two addresses, in particular the department within the organization, may be different.

---

[1] This idea is not applicable in fields in which the authors of a publication tend to be listed in alphabetical order. We refer to Frandsen and Nicolaisen (2010) and Waltman (2012) for detailed analyses of the phenomenon of alphabetical authorship.

Table 1. The authors of our example publication and the corresponding addresses.

| Author | Address |
|---|---|
| Author 1 (first author) | Address 1 |
| Author 2 | Address 1; Address 2 |
| Author 3 | Address 3 |
| Author 4 (corresponding author) | Address 3 |
| Author 5 | Address 4; Address 5 |

Table 2. The addresses of our example publication and the corresponding organizations and countries.

| Address | Organization | Country |
|---|---|---|
| Address 1 | Organization 1 | Country 1 |
| Address 2 | Organization 1 | Country 1 |
| Address 3 | Organization 2 | Country 2 |
| Address 4 | Organization 3 | Country 2 |
| Address 5 | Organization 4 | Country 3 |

Three important units of analysis in bibliometric studies are authors, organizations, and countries. We therefore look at our example from the point of view of each of these three units of analysis. We start by taking authors as the unit of analysis. Table 3 shows for different counting methods the weight with which our example publication is assigned to each of the five authors. When authors are the unit of analysis, only one of the four fractional counting variants discussed above needs to be taken into consideration. Address-level, organization-level, and country-level fractional counting are of little interest in this situation, and therefore only author-level fractional counting is included in Table 3. The table shows that in the full counting case each author has a weight of one, while in the fractional counting case the authors each have a weight of $1 / 5 = 0.20$. In the case of first author counting, author 1 has a weight of one and the other authors have a weight of zero. Author 4 is the corresponding author of our example publication (see Table 1), and therefore this author has a weight of one in the case of corresponding author counting while the other authors have a weight of zero.

Table 3. The weights with which our example publication is assigned to the five authors. The weights are presented for four different counting methods.

|  | Author 1 | Author 2 | Author 3 | Author 4 | Author 5 |
|---|---|---|---|---|---|
| Full counting | 1.00 | 1.00 | 1.00 | 1.00 | 1.00 |
| Author-level fractional counting | 0.20 | 0.20 | 0.20 | 0.20 | 0.20 |
| First author counting | 1.00 | 0.00 | 0.00 | 0.00 | 0.00 |
| Corresponding author counting | 0.00 | 0.00 | 0.00 | 1.00 | 0.00 |

We now focus on organizations as the unit of analysis. This gives the results that are reported in Table 4. Three fractional counting variants are taken into consideration, namely author-level, address-level, and organization-level fractional counting. Country-level fractional counting is not included, because it is of little interest when organizations are the unit of analysis. The results obtained using full, first author, and corresponding author counting require no further explanation, so we focus on the fractional counting results. In the case of organization-level fractional counting, each organization has equal weight. There are four organizations, which means that each organization has a weight of 1 / 4 = 0.25. Address-level fractional counting gives equal weight to each of the five addresses listed in the address list of our example publication. So each address has a weight of 1 / 5 = 0.20. Since organization 1 is mentioned in two addresses (see Table 2), this organization has a weight of 2 × 0.20 = 0.40. The other three organizations are each mentioned in only one address, and these organizations therefore each have a weight of 0.20. In the case of author-level fractional counting, each author is given an equal weight of 1 / 5 = 0.20. Based on Tables 1 and 2, it can be seen that two authors (i.e., authors 1 and 2) are affiliated to organization 1 while another two authors (i.e., authors 3 and 4) are affiliated to organization 2. Organizations 1 and 2 therefore each have a weight of 2 × 0.20 = 0.40. There is only one author who is affiliated to organizations 3 and 4, and for both organizations this is the same author (i.e., author 5). The weight of this author therefore needs to be shared by the two organizations. For this reason, organizations 3 and 4 each have a weight of 0.20 / 2 = 0.10.

Table 4. The weights with which our example publication is assigned to the four co-authoring organizations. The weights are presented for six different counting methods.

|  | Org. 1 | Org. 2 | Org. 3 | Org. 4 |
|---|---|---|---|---|
| Full counting | 1.00 | 1.00 | 1.00 | 1.00 |
| Organization-level fractional counting | 0.25 | 0.25 | 0.25 | 0.25 |
| Address-level fractional counting | 0.40 | 0.20 | 0.20 | 0.20 |
| Author-level fractional counting | 0.40 | 0.40 | 0.10 | 0.10 |
| First author counting | 1.00 | 0.00 | 0.00 | 0.00 |
| Corresponding author counting | 0.00 | 1.00 | 0.00 | 0.00 |

Finally, we consider countries as the unit of analysis. The results are presented in Table 5. All four fractional counting variants are included. The results in Table 5 have been obtained in a similar way as the results in Tables 3 and 4. We therefore do not discuss these results in more detail.

Table 5. The weights with which our example publication is assigned to the three co-authoring countries. The weights are presented for seven different counting methods.

|  | Country 1 | Country 2 | Country 3 |
|---|---|---|---|
| Full counting | 1.00 | 1.00 | 1.00 |
| Country-level fractional counting | 0.33 | 0.33 | 0.33 |
| Organization-level fractional counting | 0.25 | 0.50 | 0.25 |
| Address-level fractional counting | 0.40 | 0.40 | 0.20 |
| Author-level fractional counting | 0.40 | 0.50 | 0.10 |
| First author counting | 1.00 | 0.00 | 0.00 |
| Corresponding author counting | 0.00 | 1.00 | 0.00 |

**2.3. Choosing between different fractional counting variants**

In the above example, we have seen that there are three or four different fractional counting variants that can be used in a bibliometric study with organizations or countries as the unit of analysis. It may seem a natural choice to use organization-level fractional counting when organizations are the unit of analysis and country-level fractional counting when countries are the unit of analysis. However, a number of arguments can be provided to make clear that this is not always the best choice. Below we give three arguments. The first two arguments are of a practical nature. The third argument is more fundamental.

The first argument is that a bibliometric study sometimes involves multiple units of analysis. In order to have consistent results for the different units of analysis, it may then be necessary to calculate all results using the same fractional counting variant. For instance, in a study that involves both organizations and countries, one may want to ensure consistent results by calculating all results using organization-level fractional counting (or author-level or address-level fractional counting), not only the results for organizations but also the results for countries.

The second argument relates specifically to bibliometric studies in which organizations are the unit of analysis. Bibliographic databases such as Web of Science and Scopus contain the names of organizations as reported by the authors of a publication. However, the way in which authors report the names of organizations often does not correspond with the way in which organizations are defined for the purpose of a bibliometric study. For instance, the authors of a publication may report two organization names, 'Leiden University' and 'Leiden University Medical Center', while for the purpose of a bibliometric study both organization names may be considered to represent the same organization. In many cases, organization names need to be unified in order to obtain a proper match with the organizational definitions used in a bibliometric study. However, if we want to apply organization-level fractional counting in a consistent way, a unification needs to be performed not only for the names of the organizations included in a bibliometric study but also for the names of all co-authoring organizations. Performing such a comprehensive unification of organization names can be extremely time consuming and may therefore not feasible. In that case, we cannot apply organization-level fractional counting in a consistent way. As an alternative, either address-level or author-level fractional counting can be used, since these fractional counting variants do not require a comprehensive unification of organization names. An example of a bibliometric analysis in which organization-level fractional counting cannot be used is the CWTS Leiden Ranking. This is a ranking of 750 major universities worldwide using citation-based indicators (see www.leidenranking.com; Waltman et al., 2012). Instead of organization-level fractional counting, the CWTS Leiden Ranking uses address-level fractional counting.

There is a more fundamental argument why author-level or address-level fractional counting may be considered preferable over organization-level or country-level fractional counting. Suppose we have a publication with ten authors, of which nine are affiliated with organization 1 and one is affiliated with organization 2. Organization-level fractional counting assigns this publication to each organization with a weight of 0.5. Author-level fractional counting, on the other hand, assigns the publication with a weight of 0.9 to organization 1 and with a weight of 0.1 to organization 2. Without further information, we do not know which of the two fractional counting variants better reflects the contributions made by the two organizations. Nevertheless, in the case of a publication like this one, it seems reasonable to assume that in general the contribution made by organization 1 is substantially larger than the contribution made by organization 2. In that case, author-level fractional counting would better reflect the contributions of the two organizations than organization-level fractional counting, and therefore author-level fractional counting would be the preferred counting method. A similar argument may be given for preferring address-level fractional counting over organization-level fractional counting. In addition, one may use a similar reasoning to argue against country-level fractional counting and for author-level or address-level fractional counting.

**2.4. Earlier work on counting methods**

Important work on counting methods is reported by Gauffriau, Larsen, and colleagues (Gauffriau & Larsen, 2005; Gauffriau, Larsen, Maye, Roulin-Perriard, & Von Ins, 2007, 2008; Larsen, 2008). Gauffriau et al. (2007) propose a systematic terminology for counting methods. What we call full counting in this paper is referred to as whole counting in the proposal of Gauffriau et al. (2007),[2] while first author counting is referred to as straight counting and fractional counting as normalized counting. A further distinction is made between whole-normalized counting and complete-normalized counting. Whole-normalized counting describes the situation in which each co-author of a publication is given the same weight. Complete-normalized counting refers to the situation in which some co-authors may be given more weight than others. For instance, when countries are the unit of analysis, country-level fractional counting would be referred to as whole-normalized counting while author-level, address-level, and organization-level fractional counting would be referred to as complete-normalized counting.

---

[2] In addition to whole counting, full counting is also sometimes referred to as integer counting or total counting in the literature.

Gauffriau and Larsen (2005) and Gauffriau et al. (2008) present empirical comparisons of counting methods at the level of countries. Their focus is on publication output, not on citation impact. Gauffriau et al. (2008) also provide an extensive overview of earlier literature on counting methods. Larsen (2008) investigates the use of counting methods in publications by bibliometric researchers.

Moed (2005) also compares counting methods at the level of countries, focusing on publication output. He emphasizes that each counting method provides different information, and he therefore suggests the combined use of multiple counting methods.

Recent work on the comparison of counting methods is reported by Huang, Lin, and Chen (2011) and Lin, Huang, and Chen (2013). These authors take into account both publication output and citation impact. Because their focus is on a single field of science (i.e., physics), they do not consider the issue of field normalization. Huang, Lin, and Chen (2011) present an analysis at the level of countries, while the analysis of Lin, Huang, and Chen (2013) takes place at the level of organizations (i.e., universities). In both analyses, the authors indicate that they prefer fractional and first author counting over full counting.

Aksnes, Schneider, and Gunnarsson (2012) compare full and fractional counting at the level of countries. Their focus is on field-normalized citation-based indicators. They conclude that there are strong arguments in favor of fractional counting.

Waltman et al. (2012) present a comparison of full and fractional counting at the level of organizations. Their analysis considers the 500 universities included in the 2011/2012 edition of the CWTS Leiden Ranking. Like Aksnes et al. (2012), Waltman et al. focus on field-normalized citation-based indicators. Waltman et al. argue that at the level of organizations fractional counting is preferable over full counting. In the present paper, we extend the work of Waltman et al. by elaborating the argument in favor of fractional counting in more detail and by providing an extensive empirical analysis.

At the level of authors rather than organizations and countries, discussions on counting methods go back to Lindsey (1980) and De Solla Price (1981). De Solla Price argues that at the author level fractional counting is preferable over full counting. Lindsey presents an overview of author-level bibliometric analyses in the sociology of science literature. Most analyses turn out to use full counting, but Lindsey recommends the use of fractional counting. The introduction of the h-index (Hirsch, 2005) has led to a renewed interested in counting methods at the author level. Fractional counting variants of the h-index are studied by Egghe (2008) and Schreiber (2008a, 2008b, 2009).

Finally, in order to avoid possible confusion, we note that the term 'fractional counting' is also used in a number of recent papers by Leydesdorff and colleagues (e.g., Leydesdorff & Opthof, 2010). In these papers, fractional counting refers to a method for field normalization of citation-based indicators. It does not refer to a method for fractionally assigning publications to co-authors. The work by Leydesdorff and colleagues therefore has no direct relevance in the context of the present paper.

## 3. Relation between counting methods and field normalization

Our aim in this section is to demonstrate the close connection between counting methods and field normalization. In particular, we aim to make clear that full counting is fundamentally inconsistent with the idea of field normalization. We argue that full counting yields results that suffer from a bias in favor of fields in which there is a lot of co-authorship and in which co-authorship correlates with additional citations. This bias is caused by the fact that co-authored publications are counted multiple times in the case of full counting, once for each co-author.

We present our argument by providing two simple examples. Both examples take countries as the unit of analysis, and they both focus on the comparison between full counting and country-level fractional counting. However, the underlying ideas of the two examples are more general, and similar examples can be given with authors or organizations as the unit of analysis and with other fractional counting variants. In both examples, field normalization is performed using the mean normalized citation score (MNCS) indicator (Waltman, Van Eck, Van Leeuwen, Visser, & Van Raan, 2011; see also Lundberg, 2007). Again, the underlying ideas are more general, and similar examples can be given using other field-normalized indicators.

### 3.1. Example involving a single field

We consider a world in which there are just four publications. These publications have been produced by two countries, labeled as country A and country B. Table 6 shows for each publication the countries by which the publication is authored and the number of citations the publication has received. The table also shows the normalized citation score of each publication. For simplicity, it is assumed that all four publications are in the same field. The normalized citation score of a publication is therefore obtained simply by dividing the number of citations of the publication by the average number of citations of all four publications. The average number of citations of the four publications equals (3 + 6 + 1 + 10) / 4 = 5, and therefore the normalized citation score of for instance publication 1 equals 3 / 5 = 0.6. Of course, the average of the normalized citation scores of the four publications equals one.

Table 6. Example involving a single field.

|  | Authors | No. of cit. | Norm. cit. score |
|---|---|---|---|
| Publication 1 | Country A | 3 | 0.6 |
| Publication 2 | Country A | 6 | 1.2 |
| Publication 3 | Country B | 1 | 0.2 |
| Publication 4 | Country A; Country B | 10 | 2.0 |

We now calculate both for country A and for country B the MNCS. Using full counting, we obtain

$$\text{MNCS}_A = \frac{0.6 + 1.2 + 2.0}{3} = 1.27 \qquad (1)$$

and

$$\text{MNCS}_B = \frac{0.2 + 2.0}{2} = 1.10. \qquad (2)$$

On the other hand, using fractional counting, we get

$$\text{MNCS}_A = \frac{1.0 \times 0.6 + 1.0 \times 1.2 + 0.5 \times 2.0}{1.0 + 1.0 + 0.5} = 1.12 \qquad (3)$$

and

$$\mathrm{MNCS_B} = \frac{1.0 \times 0.2 + 0.5 \times 2.0}{1.0 + 0.5} = 0.80, \qquad (4)$$

where publication 4 has been assigned with a weight of 0.5 to country A and with a weight of 0.5 to country B.

The important thing to observe in this example is that in the case of full counting country A and country B both have an MNCS above one. One of the main ideas of field-normalized indicators such as the MNCS indicator is that the value of one can be interpreted as the world average. Under this interpretation, country A and country B both perform above the world average. Since there are no other countries in our example, the conclusion would be that all countries in the world perform above the world average. There are no countries with a below-average performance. In our opinion, the conclusion that everyone is above average does not make much sense. Moreover, this conclusion is fundamentally different from the conclusion that is reached in the case of fractional counting. Using fractional counting, country A has a performance above the world average while the performance of country B is below the world average.

Looking a bit more in detail at our example, we observe that in the fractional counting case we have

$$\frac{2.5 \times \mathrm{MNCS_A} + 1.5 \times \mathrm{MNCS_B}}{2.5 + 1.5} = \frac{2.5 \times 1.12 + 1.5 \times 0.80}{2.5 + 1.5} = 1. \qquad (5)$$

Hence, the weighted average of the MNCS of country A and the MNCS of country B, with weights given by each country's fractional number of publications, equals exactly one. This is a general property of fractional counting. The weighted average of the MNCSs of all countries in the world will always be equal to exactly one.

In the full counting case, the weighted average of the MNCS of country A and the MNCS of country B equals

$$\frac{3 \times \text{MNCS}_A + 2 \times \text{MNCS}_B}{3 + 2} = \frac{3 \times 1.27 + 2 \times 1.10}{3 + 2} = 1.20, \tag{6}$$

where the weight of each country is given by the number of publications of the country obtained using full counting. Eq. (6) tells us that in the full counting case the world average at the country level does not equal one but instead equals 1.20. Taking 1.20 as the world average, we conclude that country A, with an MNCS of 1.27, has an above-average performance while country B, with an MNCS of 1.10, performs below average. This is in agreement with the conclusion reached using fractional counting.

So in our example there is a difference of 1.20 – 1 = 0.20 between the world average obtained using full counting and the world average obtained using fractional counting. We refer to this difference as the full counting bonus. In principle, the full counting bonus can be either positive or negative, but in Section 4 we will see that in practice the bonus is usually positive. The full counting bonus is caused by the fact that publications co-authored by multiple countries are counted multiple times in the case of full counting, and therefore the citation impact of multi-country publications relative to single-country publications determines whether the full counting bonus is positive or negative. The bonus will be positive if publications co-authored by multiple countries receive more citations than publications authored by a single country. Conversely, a negative bonus will be obtained if multi-country publications are cited less frequently than single-country publications. As can be seen in Table 6, in our example the only publication co-authored by multiple countries is publication 4, and this is also the most highly cited publication. In the full counting case, publication 4 is fully assigned both to country A and to country B. Hence, the most highly cited publication in our example is counted two times, once for country A and once for country B. This double counting of publication 4 explains why both countries have an MNCS above one and why the full counting bonus is positive.

**3.2. Example involving multiple fields**

In the example discussed above, all publications are in the same field. We now consider an example that involves more than one field. This example is presented in Table 7. There are six publications, three in field X and three in field Y, and there are four countries. Countries A and B are active only in field X, while countries C and D are active only in field Y. The three publications in field X have all received the same number of citations, and therefore these publications all have a normalized citation score of one. This is not the case in field Y, in which publication 6, co-authored by countries C and D, has received more citations than publications 4 and 5, which are single-country publications. Of course, the average normalized citation score of the publications in field Y equals one, just like in field X.

Table 7. Example involving multiple fields.

|  | Field | Authors | No. of cit. | Norm. cit. score |
|---|---|---|---|---|
| Publication 1 | Field X | Country A | 10 | 1.0 |
| Publication 2 | Field X | Country B | 10 | 1.0 |
| Publication 3 | Field X | Country A; Country B | 10 | 1.0 |
| Publication 4 | Field Y | Country C | 4 | 0.8 |
| Publication 5 | Field Y | Country D | 4 | 0.8 |
| Publication 6 | Field Y | Country C; Country D | 7 | 1.4 |

Using fractional counting, the four countries all have an MNCS of exactly one. For countries A and B this is immediately clear. In the case of countries C and D, the MNCS is calculated as $(1.0 \times 0.8 + 0.5 \times 1.4) / (1.0 + 0.5) = 1$. So fractional counting tells us that all four countries perform at the world average. This is indeed the outcome that we would expect to obtain. The publications of countries A and B have all been cited equally frequently as the average of their field, so countries A and B obviously perform at the world average. In the case of countries C and D, we observe that these countries have exactly the same performance and that they are the only countries active in field Y. Based on these two observations, it is natural to conclude that the performance of countries C and D is at the world average.

We now consider the full counting case. Using full counting, countries A and B have an MNCS of one, while countries C and D have an MNCS of $(0.8 + 1.4) / 2 = 1.10$. The full counting results seem to suggest that countries C and D have a better performance than countries A and B. However, a more careful analysis shows that this is not a correct interpretation of the results. To see this, we calculate both for field X and for field Y the average of the MNCSs of the countries active in the field. (We calculate simple unweighted averages because all countries have the same number of publications.) The average MNCS of the countries active in field X equals one, while the average MNCS of the countries active in field Y equals 1.10. Hence, both countries A and B active in field X and countries C and D active in field Y perform at the world average of their field. Like in the fractional counting case, we conclude that all four countries have an average performance. Countries C and D have a higher MNCS than countries A and B only because they are active in a field with a higher full counting bonus. Field Y has a full counting bonus of $1.10 - 1 = 0.10$, while the full counting bonus in field X equals zero.

**3.3. Conclusions based on the examples**

Based on the examples presented in Subsections 3.1 and 3.2, two important conclusions can be drawn. The first conclusion is that there is a need to carefully distinguish between two field normalization concepts. We refer to these concepts as weak field normalization and strong field normalization. Weak field normalization requires the average of the normalized citation scores of all publications in a field to be equal to one. Strong field normalization is more demanding. It requires the weighted average of the MNCSs of all countries active in a field to be equal to one, where the weight of a country is given by its number of publications in the field.

As shown in the above examples, full counting yields results that are in agreement with the idea of weak field normalization, but these results may violate the idea of strong field normalization. For instance, in the example discussed in Subsection 3.1, the average normalized citation score of the four publications equals one (weak field normalization), but the average MNCS of the two countries does not equal one (no strong field normalization). Fractional counting results, on the other hand, satisfy not only the idea of weak field normalization but also the idea of strong field normalization. Using fractional counting, the weighted average of the MNCSs of all countries active in a field will always be equal to one.

When citation-based indicators are calculated using full counting, there is a risk of misinterpretation. People may confuse the concepts of weak and strong field normalization, and they may fail to understand that the idea of strong field normalization does not apply in the case of full counting. In the example presented in Subsection 3.2, they may for instance draw the incorrect conclusion that countries C and D perform above the world average. In the fractional counting case, people will not draw such an incorrect conclusion, because fractional counting results are in agreement with the idea of strong field normalization.

We now turn to the second conclusion that follows from our examples. The fact that full counting yields results that are incompatible with the idea of strong field normalization may in itself be regarded as just a minor issue. Instead of having a world average of one, the average of all countries in the world may for instance be equal to 1.10 or 1.20. Although a world average of one might be somewhat more convenient, the exact value of the world average may in the end seem to be of limited importance.

However, our second conclusion is that deviations of the world average from one actually do have serious consequences, at least when making comparisons between fields. This is what is shown in the example given in Subsection 3.2. Using full counting, the average MNCS of the countries active in field X equals one, while the average MNCS of the countries active in field Y equals 1.10. So in field X the world average equals one, while in field Y we have a world average of 1.10. Direct comparisons of the MNCSs of the countries active in field X and the countries active in field Y therefore do not yield valid conclusions. Based on their MNCSs, the countries active in field Y seem to perform better than the countries active in field X, but taking into account the fact that field Y has a higher world average than field X, it actually should be concluded that all countries perform at the same level.

Essentially, the second conclusion that we draw based on our examples is that full counting is fundamentally inconsistent with the idea of field normalization. Citation-based indicators calculated using full counting yield results that do not allow for valid comparisons between fields, and this is the case even when field-normalized indicators, such as the MNCS indicator, are used. When full counting is used in the calculation of field-normalized indicators, countries that focus their activity on fields with a high full counting bonus have an advantage over countries that are active mainly in fields with a low full counting bonus. Fractional counting does not suffer from this problem. Fractional counting results are compatible with the idea of strong field normalization, and these results therefore do allow for proper between-field comparisons.

## 4. Empirical analysis of the full counting bonus

In the previous section, we have introduced the idea of the full counting bonus and we have illustrated this idea using theoretical examples. In this section, we present a large-scale empirical analysis of the full counting bonus. This analysis for instance makes clear which fields benefit most from the full counting bonus, and the analysis shows the differences between fields caused by the bonus.

**4.1. Calculation of the full counting bonus**

We first explain in more detail the way in which we calculate the full counting bonus. For simplicity, we assume that our interest is in the full counting bonus at the level of countries. However, the full counting bonus can be calculated in a similar way at the level of for instance authors or organizations.

Suppose we have a set of *n* publications. This could be for instance the set of all publications in a specific field and in a specific year. For each publication *i*, we have a citation score $c_i$. The citation score of a publication can be defined in different ways. It may be simply the number of times a publication has been cited, but it may also be something more advanced, for instance a field-normalized citation score. We also know for each publication the countries by which the publication has been co-authored. We use $m_i$ to denote the number of countries that have co-authored publication *i*.

In order to obtain the full counting bonus, we first calculate for each country the average citation score of its publications. We perform this calculation both using full counting and using fractional counting. Next, we calculate a weighted average of the average citation scores of all countries. In the case of full counting, we use the number of publications of a country obtained using full counting as the weight of the country. In the case of fractional counting, we use a country's number of publications obtained using fractional counting as the country's weight. Finally, we calculate the full counting bonus as the difference between the weighted average in the full counting case and the weighted average in the fractional counting case.

The above approach to calculating the full counting bonus is somewhat complicated. However, a mathematically equivalent but much simpler approach is available. In this approach, the full counting bonus is calculated as

$$\text{FCB} = \frac{\sum_{i=1}^{n} m_i c_i}{\sum_{i=1}^{n} m_i} - \frac{\sum_{i=1}^{n} c_i}{n}, \tag{7}$$

where the first term equals the above-mentioned weighted average in the full counting case while the second term equals the weighted average in the fractional counting case. In the first term, the citation score $c_i$ of publication *i* co-authored by $m_i$ countries is counted $m_i$ times. This is because in the full counting case publication *i* is fully assigned to each of the $m_i$ countries. In the second term, the citation score $c_i$ of publication $i$ is counted only once, regardless of the number of countries $m_i$ by which publication $i$ has been co-authored. This is because in the fractional counting case the total weight with which publication $i$ is assigned to the $m_i$ countries equals one.[3] Importantly, which fractional counting variant is considered does not make any difference, since all fractional counting variants have the property that they assign each publication with a total weight of one to the co-authoring countries. In fact, this property is also shared by first author and corresponding author counting, and therefore the second term in (7) can also be considered to represent these counting methods.

In our empirical analysis, we consider two definitions of the citation score of a publication. Both definitions include a normalization for field. In the first definition, the citation score of a publication is obtained by dividing the number of citations of the publication by the average number of citations of all publications in the same field and in the same year. Averaging the citation scores of multiple publications then gives us the MNCS indicator. This indicator was also used in the theoretical examples presented in the previous section. In the second definition of the citation score of a publication, we determine whether a publication belongs to the top 10% most frequently cited publications of its field and publication year. A publication belonging to the top 10% has a citation score of one, while a publication belonging to the bottom 90% has a citation score of zero.[4] When this second definition is used, averaging the citation scores of multiple publications yields the $PP_{top\ 10\%}$ indicator, where $PP_{top\ 10\%}$ stands for the proportion of top 10% publications (Waltman et al., 2012; Waltman & Schreiber, 2013). When the full counting bonus is calculated for the set of all publications in a specific field and in a specific year, the second term in (7) will be equal to one in the case of our first definition of the citation score of a publication. This term will be equal to 0.1 (or 10%) in the case of our second definition.[5]

---

[3] In practice, things may be somewhat more complicated. The second term in (7) is based on the assumption that each publication can be assigned to at least one country. So it is assumed that $m_i \geq 1$ for each publication $i$. In our empirical analysis, there turn out to be publications without address information. These publications cannot be assigned to any country. To calculate the full counting bonus in a proper way, these publications are left out of both the numerator and the denominator in the second term in (7).

[4] Some publications may be exactly at the boundary between the top 10% and the bottom 90%. These publications are considered to belong partially to the top 10% and partially to the bottom 90%, and therefore these publications have a citation score somewhere in between zero and one. We refer to Waltman and Schreiber (2013) for more details.

**4.2. Empirical results**

We perform our analysis using the Web of Science (WoS) database. Unless indicated otherwise, the analysis is based on publications in the period 2009–2010. Only publications of the WoS document types 'article' and 'review' are taken into account. A four-year citation window is used, including the year in which a publication appeared. For instance, in the case of a publication from 2010, citations are counted until the end of 2013. For the purpose of the calculation of the field-normalized citation scores of publications, fields are defined by the WoS journal subject categories.[6]

We consider three units of analysis: Authors, organizations, and countries. In the WoS database, a distinction is made between the regular addresses of a publication and the so-called reprint address. To determine the number of organizations and the number of countries by which a publication has been co-authored, we take into account both the regular addresses of the publication and the reprint address. The number of organizations and the number of countries of a publication is obtained by counting the number of distinct organization names and the number of distinct country names mentioned in the addresses of the publication. In some cases, this means that organization names or country names that one might consider to refer to the same organization or to the same country are not counted in that way. For instance, a publication co-authored by 'Leiden University' and 'Leiden University Medical Center' is treated as a publication with two organizations rather than one. Likewise, a publication co-authored by 'England' and 'Scotland' has two countries.

The full counting bonus depends on two factors. On the one hand, it depends on the variation among publications in the number of authors, organizations, or countries. For instance, if all publications have the same number of authors, there can be no full counting bonus at the level of authors. On the other hand, the full counting bonus also depends on the relation between the number of authors, organizations, or countries of a publication and the citation score of the publication. There can for instance be no author-level full counting bonus if publications with different numbers of authors on average all have the same citation score.

---

[5] If there are publications that cannot be assigned to any country, as discussed in Footnote 3, this property does not hold exactly.

[6] A fractional counting approach is taken to handle publications belonging to multiple subject categories. For more details, we refer to Waltman et al. (2011, Section 6).

Figure 1 presents the distribution of publications based on their number of authors, organizations, and countries (for similar results, see Gazni, Sugimoto, & Didegah, 2012). Not surprisingly, the figure shows that the variation among publications in the number of authors is largest while the variation among publications in the number of countries is smallest. Figures 2 and 3 present the relation between the number of authors, organizations, and countries of a publication and the average citation score. The average citation score is given by the MNCS indicator in Figure 2 and by the $PP_{top\ 10\%}$ indicator in Figure 3. In general, an increasing relation can be observed between the number of authors, organizations, and countries of a publication and the average citation score. The relation is strongest for countries and weakest for authors. In fact, when the number of authors is between two and five, there is hardly any dependence of the average citation score of a publication on the number of authors. Publications with three or four authors on average even have a slightly lower citation score than publications with two authors. We refer to Wuchty, Jones, and Uzzi (2007) and Franceschet and Costantini (2010) for other studies in which the relation between the number of authors of a publication and the number of citations is analyzed. A study at the level of organizations is reported by Jones, Wuchty, and Uzzi (2008).

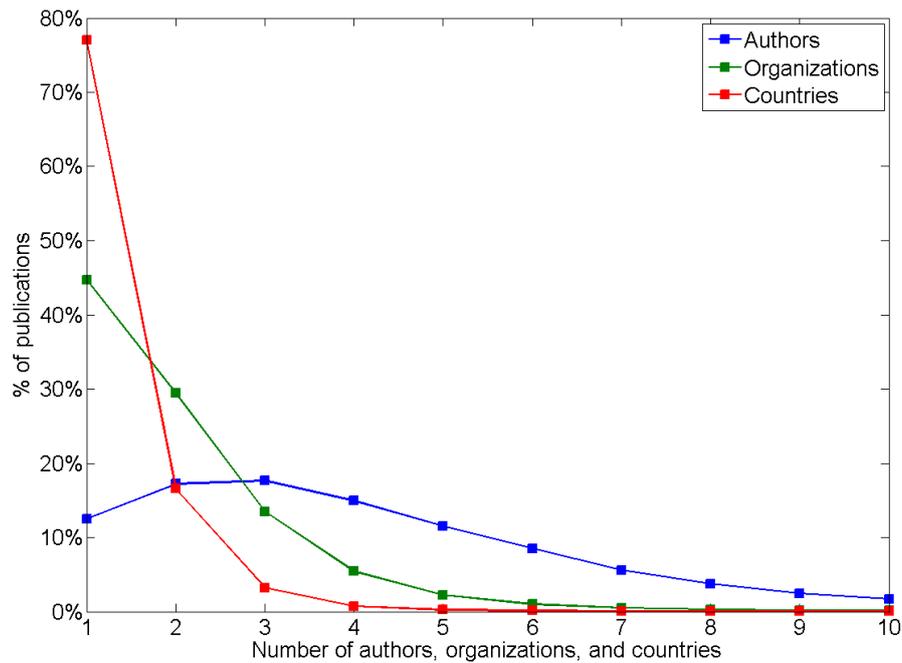

Figure 1. Distribution of publications based on their number of authors, organizations, and countries.

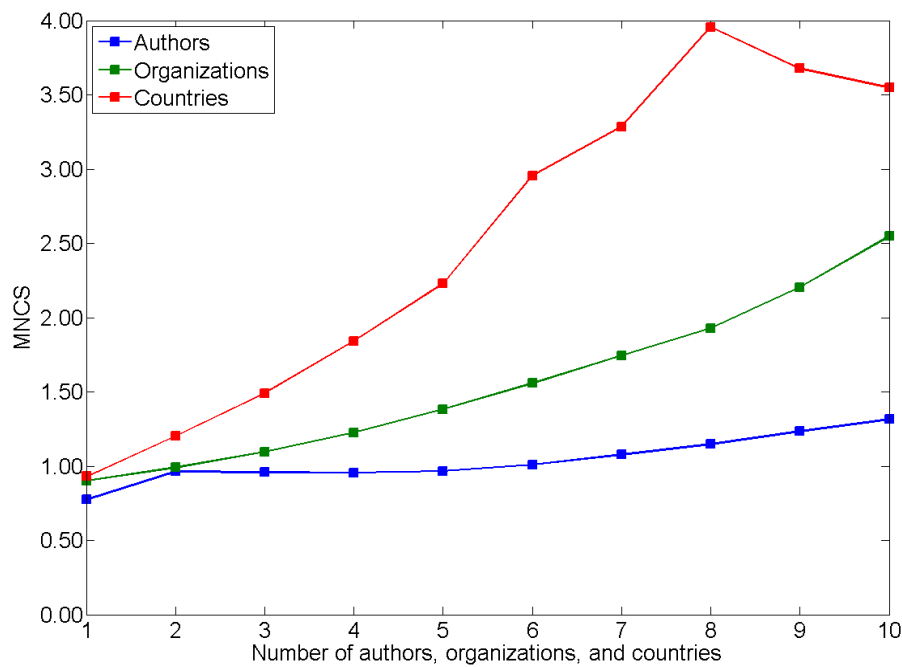

Figure 2. Relation between the number of authors, organizations, and countries of a publication and the MNCS indicator.

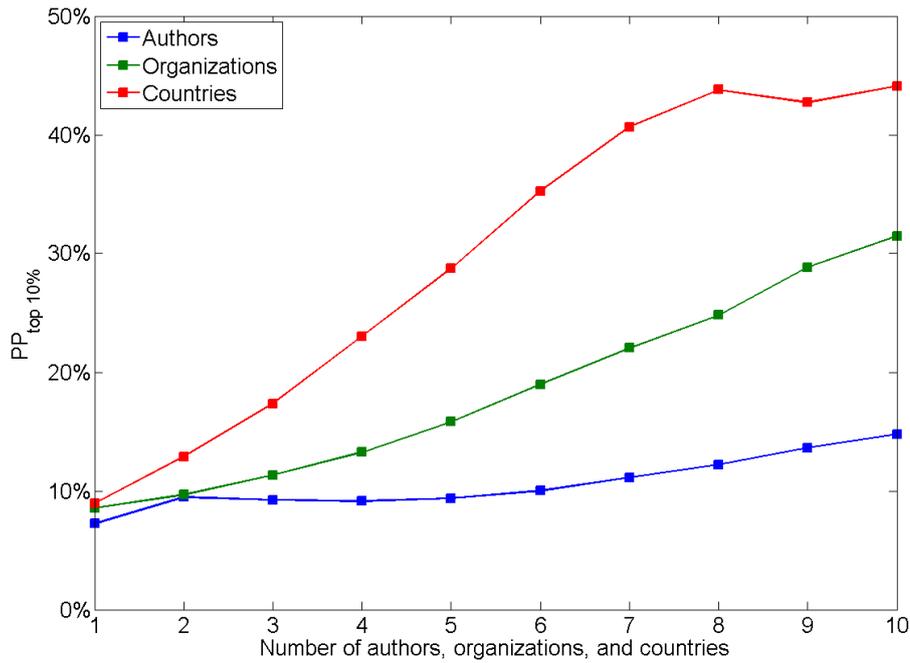

Figure 3. Relation between the number of authors, organizations, and countries of a publication and the PP$_{\text{top 10\%}}$ indicator.

Figures 1, 2, and 3 make clear that publications often have multiple co-authors and that the citation impact of a publication tends to increase with the number of co-authors. Co-authored publications are counted multiple times in the case of full counting, and our expectation based on Figures 1, 2, and 3 therefore is to observe full counting bonuses that are positive and of significant size. This is indeed what is reported in Tables 8 and 9. The tables show the full counting bonus at the level of authors, organizations, and countries for five broad fields of science and also for all fields of science taken together. Table 8 relates to the MNCS indicator, while Table 9 relates to the PP$_{\text{top 10\%}}$ indicator. In order to facilitate comparison between the results obtained for the two indicators, the full counting bonus is presented as a percentage of the average value of the indicator. For instance, in the case of the MNCS indicator, using (7) we obtain a full counting bonus of 0.248 at the level of authors for all fields of science. The average value of the MNCS indicator equals one, and therefore the full counting bonus is reported as 0.248 / 1 = 24.8% in Table 8. Likewise, the PP$_{\text{top 10\%}}$ indicator has an average value of 0.1 (or 10%), and therefore a full counting bonus of 0.0304 (or 3.04%) is reported as 0.0304 / 0.1 = 30.4% in Table 9.

Table 8. Full counting bonus for the MNCS indicator at the level of authors, organizations, and countries, including a breakdown into five broad fields of science.

|  | Authors | Organizations | Countries |
|---|---|---|---|
| All fields | 24.8% | 21.1% | 12.6% |
| Biomedical and health sciences | 20.9% | 26.8% | 16.7% |
| Life and earth sciences | 14.7% | 16.2% | 12.7% |
| Mathematics and computer science | 8.2% | 8.0% | 6.9% |
| Natural sciences and engineering | 35.2% | 19.3% | 10.8% |
| Social sciences and humanities | 14.7% | 11.2% | 5.6% |

Table 9. Full counting bonus for the $PP_{top\ 10\%}$ indicator at the level of authors, organizations, and countries, including a breakdown into five broad fields of science.

|  | Authors | Organizations | Countries |
|---|---|---|---|
| All fields | 30.4% | 26.5% | 17.1% |
| Biomedical and health sciences | 24.9% | 34.5% | 22.6% |
| Life and earth sciences | 22.8% | 24.3% | 19.7% |
| Mathematics and computer science | 11.3% | 11.3% | 9.7% |
| Natural sciences and engineering | 43.3% | 20.6% | 13.0% |
| Social sciences and humanities | 21.3% | 17.2% | 8.3% |

Based on the results for the MNCS indicator presented in Table 8, a number of conclusions can be drawn. At all three analysis levels (i.e., authors, organizations, and countries), there turns out to be a full counting bonus that is positive and of significant size. In general, the bonus is highest at the level of authors and lowest at the level of countries. We have seen in Figures 2 and 3 that the number of countries of a publication has a much stronger effect on a publication's citation score than the number of authors, but apparently this is offset by the fact that publications with a large number of countries occur much less frequently than publications with a large number of authors, as shown in Figure 1. The full counting bonus at the level of organizations is generally in between the country-level and author-level bonuses, although there are two main fields (i.e., 'Biomedical and health sciences' and 'Life and earth sciences') in which the organization-level bonus is higher than the author-level one.

The results reported in Table 8 also indicate that at the levels of authors and organizations the full counting bonus is lowest in the 'Mathematics and computer science' main field. At the country level, 'Social sciences and humanities' is the main field with the lowest bonus. The 'Natural sciences and engineering' main field has the highest bonus at the level of authors, while the highest bonus at the organization and country level can be found in the 'Biomedical and health sciences' main field.

The results for the $PP_{top\ 10\%}$ indicator reported in Table 9 are quite similar to the MNCS results presented in Table 8. However, full counting bonuses turn out to be consistently higher for the $PP_{top\ 10\%}$ indicator than for the MNCS indicator.

More detailed results at the level of 250 WoS journal subject categories can be found in an Excel file that is available at www.ludowaltman.nl/counting_methods/. The Excel file also indicates how the five main fields listed in Tables 8 and 9 are defined in terms of the WoS journal subject categories. There turn out to be rather large differences between subject categories in the full counting bonus. For instance, the subject categories with the highest MNCS full counting bonus at the level of organizations and countries are 'Medicine, general & internal' and 'Physics, nuclear'. The subject categories have bonuses of, respectively, 148% and 176% at the organization level and 89% and 70% at the country level. Other subject categories have bonuses that are close to zero or even negative. Examples of such subject categories include 'Chemistry, organic' and 'Ergonomics'.

It is important to be aware of the consequences of the large differences between subject categories in the full counting bonus. Consider a university that has a full counting MNCS of 2.50 in the 'Medicine, general & internal' subject category and a full counting MNCS of 1.00 in the 'Chemistry, organic' subject category. What should we conclude based on these values? The obvious conclusion may seem to be that in terms of citation impact our university is performing much better in the 'Medicine, general & internal' subject category than in the 'Chemistry, organic' subject category. However, this conclusion does not take into account the effect of the full counting bonus. As mentioned above, the 'Medicine, general & internal' subject category has an organization-level full counting bonus of almost 150%, while the full counting bonus for the 'Chemistry, organic' subject category is close to zero. Taking into account the effect of the full counting bonus, we need to conclude that in both subject categories our university performs around the average level of all organizations worldwide.

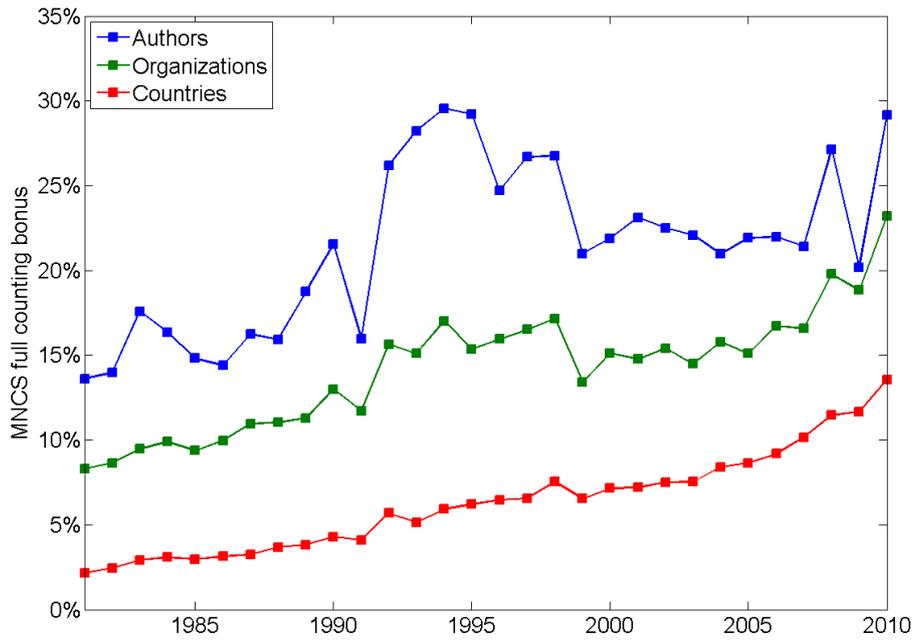

Figure 4. Development over time of the full counting bonus for the MNCS indicator at the level of authors, organizations, and countries.

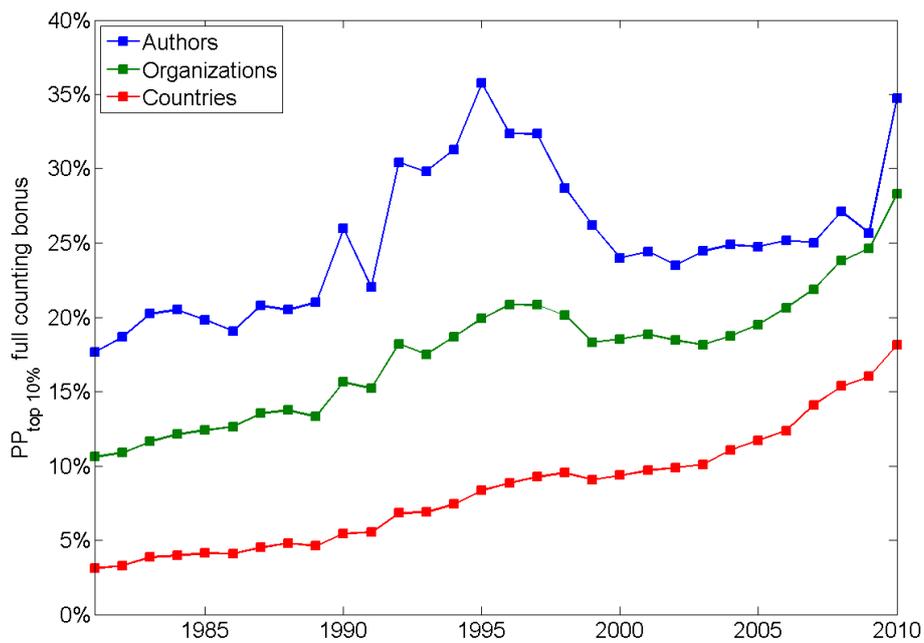

Figure 5. Development over time of the full counting bonus for the $PP_{top\ 10\%}$ indicator at the level of authors, organizations, and countries.

Finally, we look at the development of the full counting bonus over time. Figure 4 shows the development of the MNCS full counting bonus at the level of authors, organizations, and countries during the period 1981–2010. Figure 5 presents the corresponding results for the $PP_{top\ 10\%}$ indicator.[7] In general, an increasing trend can be observed. This trend is most clearly visible at the level of countries and organizations. The results at the level of authors turn out to be quite unstable. What are the implications of the increasing trend in the full counting bonus? Consider a country with a full counting MNCS of 1.02 in 1981 and a full counting MNCS of 1.14 in 2010. Taking into account the trend in the country-level full counting bonus shown in Figure 4, we conclude that, despite the increase in MNCS, both in 1981 and in 2010 the performance of our country is around the average of all countries worldwide. The citation impact of the publications co-authored by our country has increased, but the same is true for many other countries. There are more and more publications that have been co-authored by multiple countries,[8] and as shown in Figure 2, these publications tend to have a high citation impact. This causes an overall increasing trend in countries' full counting MNCSs. Because of this trend, the fact that the MNCS of a country has increased over time does not necessarily mean that the performance of the country relative to other countries has improved.

## 5. Empirical comparison of counting methods

How much difference does the choice of a counting method make in practice? To answer this question, two empirical comparisons of counting methods are presented in this section. In the first comparison, organizations are the unit of analysis. The focus of this comparison is on full counting and address-level fractional counting. The comparison aims to provide a concrete illustration of the effect of the full counting bonus. In the second comparison, countries are the unit of analysis. This comparison involves seven different counting methods.

---

[7] It is important to keep in mind that longitudinal analyses based on a database such as WoS may be influenced by changes in the coverage of the database and in the way in which the database is maintained. In the interpretation of Figures 4 and 5, the focus should therefore be on the overall time trend rather than on detailed changes over time.

[8] In 1981, 4.6% of all publications were co-authored by multiple countries. In 2010, this percentage was 21.5%.

**5.1. Comparison at the level of organizations**

Our comparison at the level of organizations is based on the 2013 edition of the CWTS Leiden Ranking.[9] This edition of the Leiden Ranking provides bibliometric statistics for 500 major universities worldwide. The 2013 edition of the Leiden Ranking is based on publications from the period 2008–2011. Our focus is on the $PP_{top\ 10\%}$ indicator. For each university, this indicator has been calculated using both full counting and address-level fractional counting. The 2013 edition of the Leiden Ranking is available online at www.leidenranking.com/ranking/2013/. A detailed discussion of the Leiden Ranking, focusing on the 2011/2012 edition, is provided by Waltman et al. (2012).

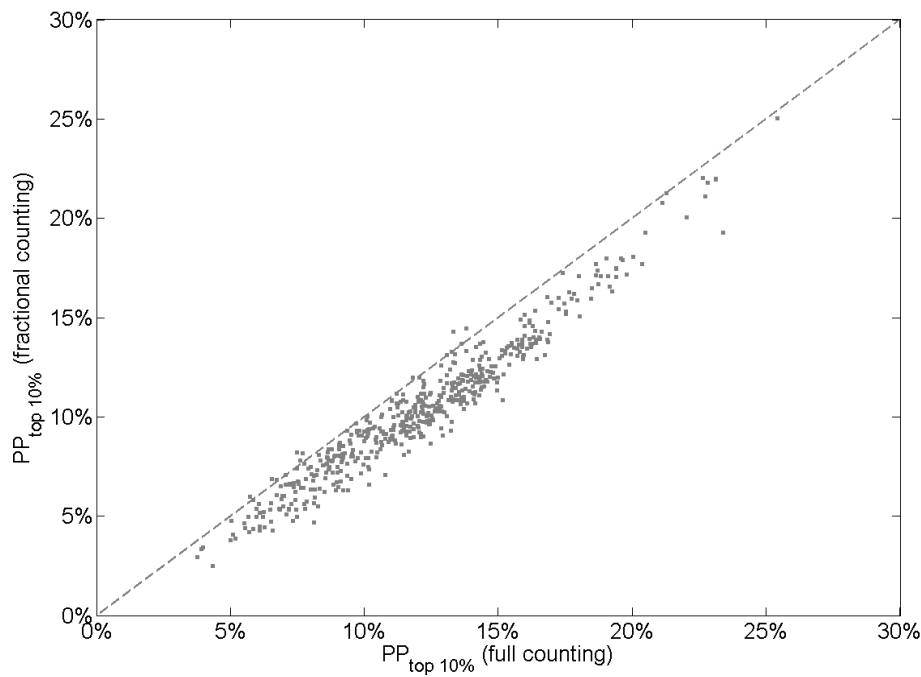

Figure 6. Scatter plot of the $PP_{top\ 10\%}$ of 500 universities according to full counting and address-level fractional counting.

Figure 6 presents a scatter plot indicating for each of our 500 universities the $PP_{top\ 10\%}$ calculated using full counting and the $PP_{top\ 10\%}$ calculated using address-level fractional counting. For almost all universities, the $PP_{top\ 10\%}$ obtained using fractional counting turns out to be lower than the PP$_{\text{top 10\%}}$ obtained using full counting. This is to be expected, given the fact that results based on full counting benefit from a full counting bonus. On average, the PP$_{\text{top 10\%}}$ of a university is almost 2 percentage points lower in the case of fractional counting than in the case of full counting.

---

[9] We use the 2013 edition of the Leiden Ranking rather than the more recent 2014 edition. This is because the 2013 edition employs the same five broad fields of science that were also considered in Section 4.

Based on Figure 6, it can be concluded that overall there is a strong correlation between the full counting PP$_{\text{top 10\%}}$ indicator and the address-level fractional counting PP$_{\text{top 10\%}}$ indicator. However, Figure 6 also shows that for some universities the difference between the two counting methods is much larger than for others. On the one hand, there are universities for which the fractional counting PP$_{\text{top 10\%}}$ is more than 4 percentage points lower than the full counting PP$_{\text{top 10\%}}$. On the other hand, we also observe universities for which the fractional counting PP$_{\text{top 10\%}}$ is approximately equal to or even somewhat higher than the full counting PP$_{\text{top 10\%}}$.

In the previous sections, we have shown that full counting may yield results that are biased in favor of certain fields of science. We now provide a concrete example of this possibility. Of the five broad fields of science considered in Section 4, 'Biomedical and health sciences' has the highest full counting bonus at the level of organizations (see Tables 8 and 9). This suggests that for universities with a large share of their publications in the 'Biomedical and health sciences' field the difference between the full counting PP$_{\text{top 10\%}}$ and the fractional counting PP$_{\text{top 10\%}}$ on average will be larger than for universities with a small share of 'Biomedical and health sciences' publications. So differences between universities in their share of 'Biomedical and health sciences' publications may to a certain degree explain why some universities have a larger difference between their full counting PP$_{\text{top 10\%}}$ and their fractional counting PP$_{\text{top 10\%}}$ than others, as shown in Figure 6.

Figure 7 provides clear evidence in support of this idea. When we classify our 500 universities into five equally-sized groups based on their share of publications in the 'Biomedical and health sciences' field, the 100 universities with the largest share of 'Biomedical and health sciences' publications on average turn out to have the largest difference between their full counting PP$_{\text{top 10\%}}$ and their fractional counting PP$_{\text{top 10\%}}$, followed by the 100 universities with the second largest share of 'Biomedical and health sciences' publications, and so on. The 100 universities with the smallest share of 'Biomedical and health sciences' publications on average have the smallest difference between their full counting PP$_{\text{top 10\%}}$ and their fractional counting PP$_{\text{top 10\%}}$. The results presented in Figure 7 are in full agreement with the idea that in the case of full counting organizations whose activity is focused on fields with a high full counting bonus have an advantage over organizations that are active mainly in fields with a low full counting bonus. Comparing the 100 universities that focus most on the 'Biomedical and health sciences' field with the 100 universities that focus least on this field, a difference of almost 1.5 percentage point can be observed for the $PP_{top\ 10\%}$ indicator.

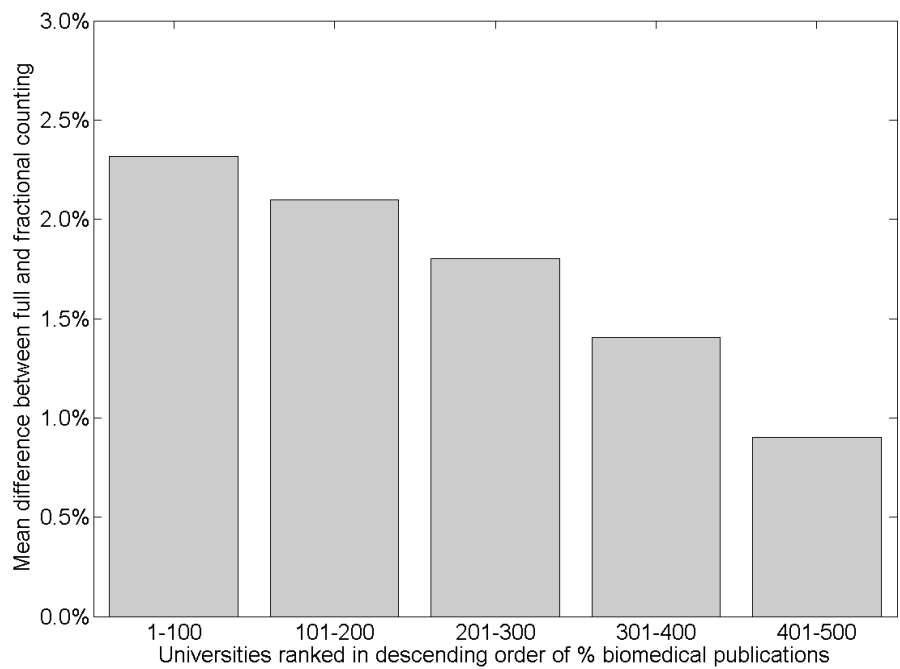

Figure 7. Average difference between the full counting $PP_{top\ 10\%}$ and the fractional counting $PP_{top\ 10\%}$ for five groups of 100 universities each. The first group includes the 100 universities with the largest share of publications in the 'Biomedical and health sciences' field, the second group includes the 100 universities with the second largest share of publications in this field, and so on.

We refer to Waltman et al. (2012) for a further analysis of the differences between full and fractional counting in the context of the CWTS Leiden Ranking. We also note that the data on which the above analysis is based can be downloaded in an Excel file from the Leiden Ranking website (www.leidenranking.com).

**5.2. Comparison at the level of countries**

We now turn to the comparison of counting methods at the level of countries. Our analysis is based on publications in the WoS database in the period 2009–2010 (considering only the document types 'article' and 'review'). We take into account the 25 countries with the largest number of publications (calculated using full counting). Both the MNCS and the $PP_{top\ 10\%}$ indicator are included in the analysis. Like in Section 4, a four-year citation window is used. We compare full counting with six alternative counting methods: Author-level fractional counting, address-level fractional counting, organization-level fractional counting, country-level fractional counting, first author counting, and corresponding author counting. Our analysis is similar to an earlier analysis reported by Aksnes et al. (2012), but this earlier analysis includes only two counting methods, namely full counting and address-level fractional counting. Also, the analysis by Aksnes et al. does not include the $PP_{top\ 10\%}$ indicator.

The detailed results of our analysis are reported in Tables A1, A2, and A3 in the appendix. Table A1 shows for each of the 25 countries included in the analysis the number of publications according to each of the seven counting methods. Tables A2 and A3 show for each country and each counting method the MNCS and the $PP_{top\ 10\%}$. The appendix also provides some technical information related to the calculation of the results presented in Tables A1, A2, and A3.

Based on the results reported in the appendix, we have selected a number of key findings of our analysis. We now discuss these findings in more detail.

Table 10 shows for the 25 countries included in our analysis the number of publications according to both full counting and address-level fractional counting. The table also shows the percentage decrease in the number of publications when we move from full counting to address-level fractional counting. As can be seen in the table, there are large differences between countries in the percentage decrease. The decrease ranges from 9% for Turkey to 42% for Switzerland and Scotland. These differences result from the fact that some countries, such as Switzerland and Scotland, are much more involved in international collaboration than other countries, such as Turkey.[10]

---

[10] Aksnes et al. (2012) observe a strong negative correlation between the size of a country in terms of publication output and the degree to which a country is involved in international collaboration. Our

Table 10. Number of publications of 25 countries according to full counting and address-level fractional counting. The fractional counting results have been rounded to integer values.

| Country | Full counting | Fractional counting | Decrease |
| --- | --- | --- | --- |
| United States | 693,107 | 580,142 | 16% |
| China | 265,850 | 232,774 | 12% |
| Germany | 179,586 | 127,200 | 29% |
| England | 163,121 | 111,057 | 32% |
| Japan | 152,216 | 130,310 | 14% |
| France | 129,845 | 91,369 | 30% |
| Canada | 112,314 | 81,626 | 27% |
| Italy | 104,383 | 78,398 | 25% |
| Spain | 90,324 | 67,672 | 25% |
| India | 82,539 | 73,117 | 11% |
| Australia | 80,042 | 58,675 | 27% |
| South Korea | 79,233 | 67,770 | 14% |
| Brazil | 64,801 | 55,784 | 14% |
| Netherlands | 62,458 | 41,976 | 33% |
| Russia | 56,048 | 45,587 | 19% |
| Taiwan | 48,710 | 42,920 | 12% |
| Turkey | 44,998 | 40,844 | 9% |
| Switzerland | 44,806 | 25,936 | 42% |
| Sweden | 40,118 | 25,773 | 36% |
| Poland | 38,886 | 30,902 | 21% |
| Belgium | 34,518 | 21,578 | 37% |
| Iran | 31,998 | 28,783 | 10% |
| Scotland | 25,553 | 14,740 | 42% |
| Israel | 24,156 | 17,682 | 27% |
| Denmark | 23,493 | 14,795 | 37% |

results do not confirm this observation. Among the ten smallest countries listed in Table 10, there are three, namely Iran, Taiwan, and Turkey, that have only a small decrease (at most 12%) in their number of publications when we move from full counting to fractional counting. So the involvement of these three countries in international collaboration is low, while at the same time these countries have a relatively small publication output. Iran, Taiwan, and Turkey are not included in the analysis performed by Aksnes et al.

Results very similar to the ones reported in Table 10 are obtained when full counting is compared not with address-level fractional counting but with any of the other alternative counting methods. In fact, it turns out that the six alternative counting methods all yield fairly similar publication statistics. The largest difference can be observed between country-level fractional counting and first author counting. As can be seen in Table A1, the number of publications of Scotland equals 15,356 according to country-level fractional counting and 14,500 according to first author counting, which is a difference of 6%. The differences between author-level, address-level, and organization-level fractional counting turn out to be especially small. For all 25 countries, the differences between these three counting methods are below 1.5%.

We now move from publication to citation statistics. Table 11 reports for each of our 25 countries the MNCS according to both full counting and address-level fractional counting. A scatter plot of the MNCSs obtained using the two counting methods is presented in Figure 8. Based on Table 11 and Figure 8, we observe that for all 25 countries address-level fractional counting yields a lower MNCS than full counting. This is in line with the results at the level of organizations discussed in Subsection 5.1 (see Figure 6). Interestingly, for some countries the difference in MNCS between full counting and address-level fractional counting is much larger than for other countries. Iran and the United States have the smallest difference. For Iran and the United States, the MNCS based on address-level fractional counting is, respectively, 0.03 and 0.04 lower than the MNCS based on full counting. The difference in MNCS between the two counting methods is largest for Belgium, Denmark, Scotland, Sweden, and Switzerland. For these countries, the fractional counting MNCS is between 0.21 and 0.25 lower than the full counting MNCS. As can be seen in Figure 8, the ranking of our 25 countries according to the MNCS indicator also changes somewhat depending on the choice of a counting method. Especially the ranking of the United States is sensitive to the counting method that is used. Based on full counting, the United States is ranked eight, with an MNCS of 1.34. Based on address-level fractional counting, on the other hand, the United States is ranked fourth, with an MNCS of 1.30.

Table 11. MNCS of 25 countries according to full counting and address-level fractional counting.

| Country | Full counting | Fractional counting | Decrease |
|---|---|---|---|
| United States | 1.34 | 1.30 | 0.04 |
| China | 0.95 | 0.88 | 0.06 |
| Germany | 1.25 | 1.09 | 0.16 |
| England | 1.42 | 1.27 | 0.15 |
| Japan | 0.87 | 0.79 | 0.08 |
| France | 1.19 | 1.02 | 0.17 |
| Canada | 1.28 | 1.12 | 0.16 |
| Italy | 1.18 | 1.00 | 0.18 |
| Spain | 1.16 | 0.99 | 0.16 |
| India | 0.74 | 0.67 | 0.07 |
| Australia | 1.30 | 1.16 | 0.15 |
| South Korea | 0.88 | 0.79 | 0.09 |
| Brazil | 0.72 | 0.62 | 0.10 |
| Netherlands | 1.54 | 1.36 | 0.19 |
| Russia | 0.50 | 0.34 | 0.16 |
| Taiwan | 0.92 | 0.86 | 0.06 |
| Turkey | 0.69 | 0.63 | 0.06 |
| Switzerland | 1.57 | 1.34 | 0.23 |
| Sweden | 1.38 | 1.14 | 0.25 |
| Poland | 0.72 | 0.55 | 0.17 |
| Belgium | 1.43 | 1.19 | 0.23 |
| Iran | 0.80 | 0.77 | 0.03 |
| Scotland | 1.50 | 1.26 | 0.24 |
| Israel | 1.16 | 0.97 | 0.19 |
| Denmark | 1.52 | 1.30 | 0.21 |

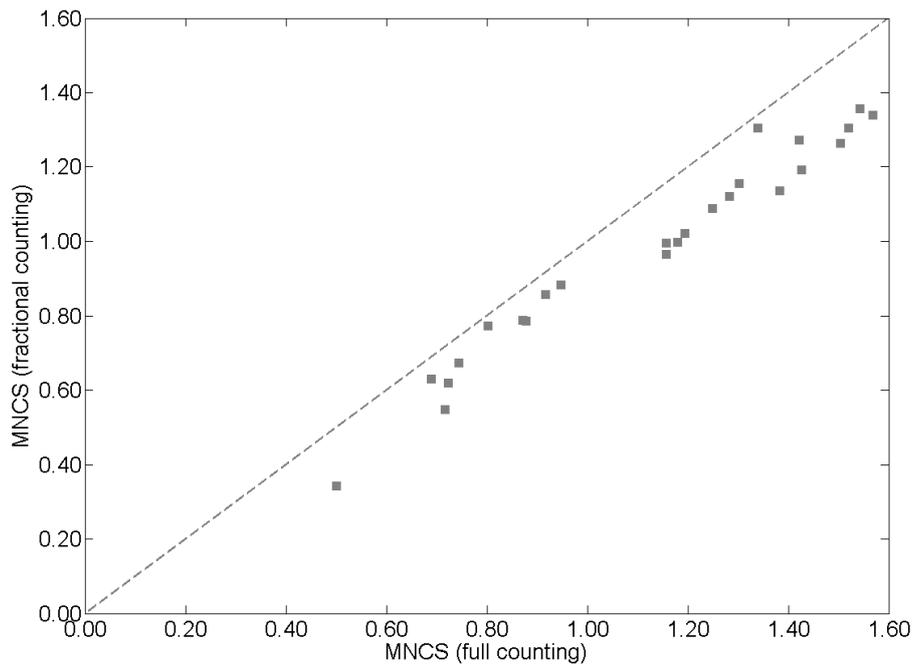

Figure 8. Scatter plot of the MNCS of 25 countries according to full counting and address-level fractional counting.

The countries with the largest decrease in their MNCS when moving from full counting to address-level fractional counting are also the countries that have the largest percentage decrease in their number of publications. On the other hand, the countries with the smallest decrease in their MNCS are countries with a small percentage decrease in their publication output. This pattern is clearly visible in Figure 9, which shows a scatter plot of countries' decrease in publication output and their decrease in MNCS. Figure 9 leads us to the conclusion that, in the case of comparisons between countries based on the MNCS indicator, full counting tends to benefit countries with a strong involvement in international collaboration while fractional counting benefits countries that are less involved in international collaboration.

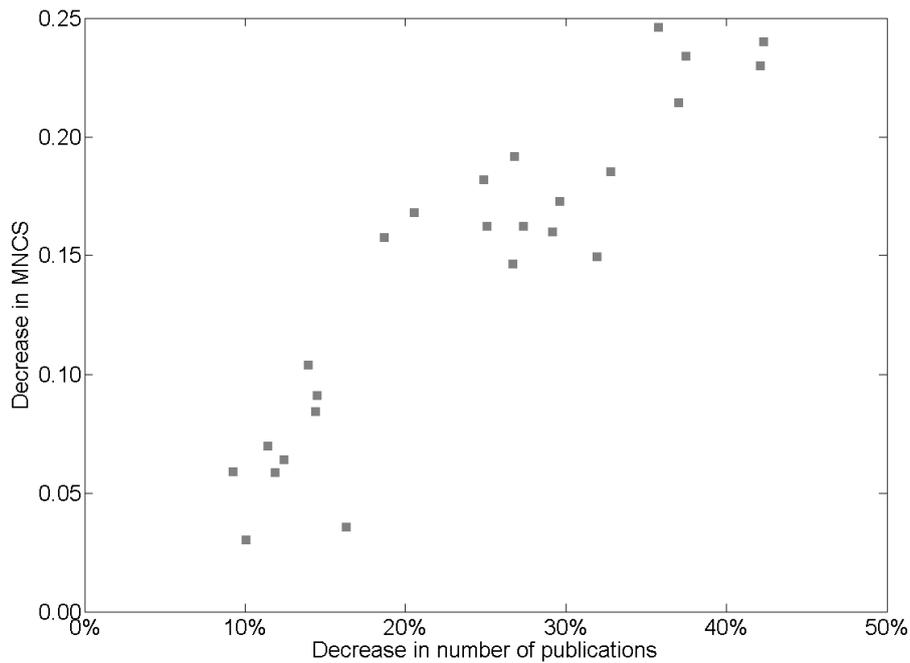

Figure 9. Scatter plot of the decrease in the number of publications of 25 countries when moving from full counting to address-level fractional counting and the corresponding decrease in the MNCS of the countries.

Instead of address-level fractional counting, we have also looked at other counting methods that can be used as an alternative to full counting. However, as can be seen in Table A2, the six alternative counting methods all yield quite similar citation statistics. The largest difference can be observed between country-level fractional counting and corresponding author counting. The MNCS of Russia equals 0.35 according to country-level fractional counting and 0.31 according to corresponding author counting. So differences in MNCS between the six alternative counting methods are at most 0.04. We note that based on the MNCS indicator Russia consistently appears as the lowest ranked country, despite the differences between the counting methods.

The effect of the choice of a counting method on the $PP_{top\ 10\%}$ indicator turns out to be very similar to the effect on the MNCS indicator. This can be seen based on the results reported in Table A3. Given the similarity between the results for the MNCS and $PP_{top\ 10\%}$ indicators, we do not discuss the $PP_{top\ 10\%}$ results in more detail.

The results of the country-level analysis presented above are generally in agreement with the results of the earlier analysis performed by Aksnes et al. (2012).

The main contribution of our analysis is to show that results at the level of countries are relatively insensitive both to the choice between different alternatives to full counting and to the choice between different citation-based indicators (i.e., MNCS and $PP_{top\ 10\%}$). We emphasize that results become more sensitive to the various choices when the number of countries included in the analysis is increased. Results for 150 instead of 25 countries are provided in an Excel file that is available at www.ludowaltman.nl/counting_methods/.

## 6. Commonly used arguments in favor of full counting

In practice, most bibliometric analyses use full counting instead of fractional counting. Below we list four arguments that are often given to argue against the use of fractional counting and to justify the use of full counting. We also provide a response to each argument.

*Argument 1: The different co-authors of a publication usually have not contributed equally. By giving equal weight to each co-author, fractional counting fails to properly represent the contributions made by the different co-authors. Hence, giving equal weight to each co-author is arbitrary and lacks a sound justification.*

It is true that there can be large differences between co-authors in the contribution they have made to a publication. At the level of an individual publication, fractional counting may therefore significantly misrepresent the contributions made by individual co-authors. However, at the level of a large set of publications, for instance all publications of an organization or a country, we believe that it is reasonable to assume that the error will be within an acceptable margin. This is because errors at the level of individual publications are likely to cancel out. The contribution of an organization or a country to certain publications may be overestimated, but most probably there will then be other publications for which the contribution of this organization or this country is underestimated.

We note that the above reasoning is somewhat similar to the reasoning that is often used to justify the use of citation analysis despite the many different reasons for which citations may be given. According to what is referred to as the 'standard account' by Nicolaisen (2007), citations may be given for various different reasons, many of which do not reflect the influence of one publication on another. However, when dealing with a sufficiently large number of citations, the effects of the different reasons for which citations are given may be expected to cancel out. It can therefore be assumed that on average citations represent the influence of one publication on another. In our opinion, if this reasoning is considered satisfactory to justify the use of citation analysis, then our above reasoning to justify the use of fractional counting should be considered acceptable too.

Furthermore, the argument that giving equal weight to each co-author of a publication is arbitrary may equally well be used as an argument against full counting. Like fractional counting, full counting gives the same weight to each co-author of a publication.

*Argument 2: Fractional counting provides an incentive against collaboration, which is often considered undesirable.*

Collaboration is indeed more attractive in the case of full counting than in the case of fractional counting, especially in fields with a significant full counting bonus. Based on this, it could be concluded that fractional counting provides an incentive against collaboration. However, an alternative viewpoint is that fractional counting is neutral with respect to collaboration while full counting provides an unfair advantage to collaboration. This unfair advantage results from the fact that collaborative publications are fully assigned to each co-author, and in many fields the advantage is reinforced by the presence of a significant full counting bonus.

More fundamentally, we believe that citation impact and collaboration represent different dimensions of scientific performance and that in general these dimensions can best be measured separately from each other. Citation-based indicators should be assessed based on the degree to which they measure citation impact in an accurate way. In this respect, we believe that for many purposes fractional counting performs better than full counting. If in addition to citation impact one also considers collaboration to be a relevant dimension of scientific performance, then additional indicators should be used to measure this dimension. If one desires to do so, these indicators can then be used to provide an incentive to collaboration. By assessing citation-based indicators based on the effect they may have on collaboration, one fails to make a proper distinction between the citation impact dimension of scientific performance and the collaboration dimension.

*Argument 3: Fractional counting is more difficult to understand and less intuitive than full counting.*

To a certain degree, we agree with this argument. Fractional counting yields non-integer publication and citation counts. These non-integer counts are more difficult to understand and require more explanation than the integer publication and citation counts provided by full counting. Fractional counting may also be less intuitive than full counting. For instance, consider a researcher who has produced some of his publications on his own while he has produced other publications with one or two co-authors. The researcher may feel that his co-authored publications are of similar importance to his oeuvre as his single-author publications. However, fractional counting gives less weight to the co-authored publications of the researcher than to his single-author publications. This is not in agreement with the feelings the researcher has about the importance of the different publications in his oeuvre, and therefore from the point of view of the researcher fractional counting can be regarded as less intuitive than full counting.

On the other hand, from a different point of view, it can also be argued that fractional counting is actually more intuitive than full counting. In Section 3, we have given two examples showing that field-normalized citation impact indicators calculated using full counting can easily be misinterpreted. Field-normalized indicators calculated using fractional counting are much more easy to interpret in a correct way. As explained in Subsection 3.3, this is because indicators based on fractional counting yield results that are compatible with the idea of strong field normalization. Unlike full counting indicators, fractional counting indicators therefore allow comparisons between fields to be performed in an easy and intuitive way. So from this point of view indicators based on fractional counting can be considered more intuitive than their full counting counterparts.

*Argument 4: Full counting and fractional counting measure different things (participation vs. contribution). Each of the two counting methods is therefore valid in its own way. Which counting method should be used in a particular analysis depends on what the analysis intends to measure.*

Suppose that countries are our unit of analysis. According to the above argument, full counting can then be used to measure the number of publications in which a country has participated. Fractional counting, on the other hand, can be used to measure the contribution that a country has made in terms of publication output, with co-authored publications counting only as a partial contribution. Likewise, full counting can be used to measure the citation impact of the publications in which a country has participated, while fractional counting can be used to measure the contribution that a country has made in terms of citation impact.

We agree that full counting can be interpreted in terms of participation and fractional counting in terms of contribution. Following this line of reasoning, both counting methods are valid in their own way. Nevertheless, we believe that the usefulness of the participation-based perspective is limited, at least when comparisons between fields need to be made. To illustrate this, we go back to the example provided in Subsection 3.2. Using full counting, countries C and D have a higher MNCS than countries A and B. Indeed, it can be concluded from this that the publications in which countries C and D participate on average have a higher field-normalized citation impact than the publications in which countries A and B participate. However, is it useful to draw such a conclusion? In a typical research assessment context, we think it is not. We believe that in practice there usually is a need to make comparisons between fields in which corrections have been made for all field-dependent factors. In the example given in Subsection 3.2, the participation-based perspective offered by full counting does not correct for the fact that countries C and D benefit from a full counting bonus. The contribution-based perspective offered by fractional counting does correct for this. We therefore believe that the latter perspective provides information that is more useful in a typical research assessment context.

## 7. Conclusions

In the bibliometric literature, various arguments have been given for and against different counting methods. Although we are sympathetic to many of the arguments in favor of fractional counting, these arguments are not the main focus of this paper. Instead, in this paper, we have presented a new perspective on the choice between different counting methods, leading to an important new argument in favor of fractional counting. Building on our earlier work (Waltman et al., 2012), this argument is based on the observation that the problem of choosing an appropriate counting method is closely connected to the problem of field normalization of citation-based indicators.

We have argued that from a field normalization point of view fractional counting is preferable over full counting. As we have shown, properly field-normalized results cannot be obtained using full counting, and field-normalized indicators calculated using full counting can easily be misinterpreted. Fractional counting does provide properly field-normalized results, and these results can be interpreted in a much more straightforward way than results obtained using full counting. Essentially, the problem of full counting is that co-authored publications are counted multiple times, once for each co-author, which creates an unfair advantage to fields with a lot of co-authorship and with a strong correlation between co-authorship and citations. For instance, the average full counting MNCS of all organizations or all countries active in these fields is significantly higher than one. On the other hand, fields in which co-authorship is less common or in which co-authorship does not correlate with citations are disadvantaged. Full counting yields results that are biased against organizations and countries whose activity is focused on these fields. Fractional counting does not suffer from this problem. In the case of fractional counting, each publication is counted only once, regardless of its number of co-authors, and this ensures that comparisons between fields can be made in an unbiased way.

### 7.1. Practical implications

What are the practical implications of the analysis presented in this paper? In our view, this depends on the level of aggregation at which a bibliometric study is performed. In the case of a study at a high aggregation level, such as the level of countries or organizations (e.g., university rankings), we consider it absolutely essential to use fractional counting instead of full counting. At this level, there is a serious risk of misinterpretation of full counting results. Moreover, we believe that arguments in favor of full counting, such as the ones discussed in Section 6, are of limited relevance at a high aggregation level.

The situation is more difficult at a low level of aggregation, for instance at the level of researchers or research groups. At this level, we believe that reasonable arguments can be given in favor of both full and fractional counting. Especially the third and fourth argument discussed in Section 6 play an important role at this level.

As pointed out in the third argument, full counting is in agreement with the intuitive idea that all publications of a researcher or a research group should be considered of equal importance. The fourth argument makes clear that full counting results can be given a valid interpretation by taking a viewpoint based on the idea of participation rather than contribution.

However, there is a more fundamental reason why the argument presented in this paper in favor of fractional counting is less relevant at a low level of aggregation. The argument depends on the connection between counting methods and field normalization, but the entire idea of field normalization may be seen as problematic at a low aggregation level. Field-normalized indicators have a limited accuracy (e.g., Leydesdorff & Bornmann, in press; Van Eck, Waltman, Van Raan, Klautz, & Peul, 2013), and it is questionable whether these indicators are sufficiently accurate for applications at a low aggregation level. If the accuracy of field-normalized indicators at a low aggregation level is considered insufficient, the argument presented in this paper in favor of fractional counting has no relevance at this level. There may of course still be other arguments in favor of fractional counting, but these arguments are not the main focus of this paper.

**7.2. Different variants of fractional counting**

In our analysis, we have made a further distinction between different variants of fractional counting. Our empirical results at the level of countries suggest that the differences between these variants are relatively small, at least in comparison with the differences between full and fractional counting. Nevertheless, we believe that in general it is best to use either the author-level or the address-level variant of fractional counting. We have presented a number of arguments in favor of these variants in Subsection 2.3. Instead of one of the fractional counting variants, it is also possible to use first author or corresponding author counting. Like fractional counting, these counting methods yield properly field-normalized results.

**7.3. Multiplicative counting**

Finally, we mention an alternative approach that can be taken to deal with the problems studied in this paper. If one desires to assign publications fully to each co-author while at the same time one also desires to obtain properly field-normalized results, one may consider the use of a multiplicative counting approach. In this approach, a publication co-authored by three countries is assigned fully to each of the countries. In addition, in the calculation of field-normalized indicators, for instance in the calculation of the average number of citations of the publications in a field, the publication is counted three times rather than just once. In this way, results are obtained that are properly field normalized, like in the case of fractional counting, while at the same time publications are assigned fully to each co-author, like in the case of full counting. Multiplicative counting is used at the country level in a number of studies by Ruiz-Castillo and colleagues (e.g., Albarrán, Crespo, Ortuño, & Ruiz-Castillo, 2010; Herranz & Ruiz-Castillo, 2012a, 2012b). In our view, the disadvantage of multiplicative counting is that publications do not all have the same weight in the calculation of field-normalized indicators. Nevertheless, we believe that multiplicative counting represents an interesting idea. Future research may focus on the comparison between fractional and multiplicative counting.

**Appendix**

In this appendix, we report the detailed results of the analysis presented in Subsection 5.1. The following abbreviations are used in Tables A1, A2, and A3:

- FUC: Full counting.
- AULFRC: Author-level fractional counting.

- ADLFRC: Address-level fractional counting.
- ORLFRC: Organization-level fractional counting.
- COLFRC: Country-level fractional counting.
- FAC: First author counting.
- CAC: Corresponding author counting.

Tables A1, A2, and A3 include results for the 25 countries with the largest number of publications (calculated using full counting). Results for 150 countries are provided in an Excel file that is available at www.ludowaltman.nl/counting_methods/.

A few technical remarks need to be made on the calculation of the results shown in Tables A1, A2, and A3. The total number of publications in the WoS database in the period 2009–2010 equals 2.46 million (considering only the document types 'article' and 'review'). Of these publications, 1.9% have no address information. Because of this, only 2.41 million publications can be assigned to one or more countries. The WoS database distinguishes between the regular addresses of a publication and the reprint address. We interpret the reprint address of a publication as the address of the corresponding author. Corresponding author counting is therefore based on the reprint address. The other counting methods are all based on the regular addresses of a publication. Some publications do not have regular addresses but do have a reprint address. In the case of these publications, all counting methods are based on the reprint address. Conversely, there are publications that do have regular addresses but that do not have a reprint address. For these publications, we assume the first author to be the corresponding author. Corresponding author counting then becomes identical to first author counting. Another difficulty is that for some publications no information is available on the relations between authors and addresses. For these publications, it is unclear which authors are affiliated to which addresses. Without this information, author-level fractional counting and first author counting cannot be implemented. We handle these publications by assuming that each author is affiliated to all addresses.

Finally, we should mention that England, Scotland, Wales, and North Ireland are seen as different countries in the WoS database. We follow the country definitions provided by the database, and therefore England and Scotland appear separately in our results. Wales and North Ireland are not included in the results presented in Tables A1, A2, and A3 because their number of publications is relatively small.

Table A1. Number of publications of 25 countries according to seven counting methods. Except for full counting, all counting methods yield non-integer results. These results have been rounded to integer values.

|  | FUC | AULFRC | ADLFRC | ORLFRC | COLFRC | FAC | CAC |
|---|---|---|---|---|---|---|---|
| United States | 693,107 | 578,332 | 580,142 | 578,749 | 575,285 | 567,958 | 576,405 |
| China | 265,850 | 234,366 | 232,774 | 232,498 | 231,744 | 241,459 | 237,329 |
| Germany | 179,586 | 128,595 | 127,200 | 127,565 | 127,348 | 126,953 | 127,078 |
| England | 163,121 | 111,143 | 111,057 | 111,614 | 112,690 | 109,487 | 110,762 |
| Japan | 152,216 | 130,596 | 130,310 | 130,510 | 129,804 | 128,868 | 128,978 |
| France | 129,845 | 90,662 | 91,369 | 91,878 | 90,676 | 88,469 | 89,381 |
| Canada | 112,314 | 81,255 | 81,626 | 81,411 | 82,662 | 80,396 | 81,958 |
| Italy | 104,383 | 78,717 | 78,398 | 78,300 | 77,408 | 79,336 | 78,870 |
| Spain | 90,324 | 68,064 | 67,672 | 67,633 | 67,380 | 68,348 | 68,836 |
| India | 82,539 | 73,490 | 73,117 | 73,177 | 73,027 | 74,388 | 73,268 |
| Australia | 80,042 | 58,424 | 58,675 | 58,725 | 58,958 | 58,021 | 58,754 |
| South Korea | 79,233 | 67,794 | 67,770 | 67,653 | 67,479 | 68,464 | 68,639 |
| Brazil | 64,801 | 56,061 | 55,784 | 55,753 | 55,437 | 56,874 | 56,189 |
| Netherlands | 62,458 | 42,153 | 41,976 | 41,914 | 42,274 | 41,998 | 41,818 |
| Russia | 56,048 | 45,555 | 45,587 | 45,636 | 45,752 | 46,278 | 45,324 |
| Taiwan | 48,710 | 42,790 | 42,920 | 42,840 | 42,589 | 43,187 | 43,234 |
| Turkey | 44,998 | 40,808 | 40,844 | 40,807 | 40,765 | 41,404 | 41,112 |
| Switzerland | 44,806 | 25,909 | 25,936 | 26,138 | 26,909 | 25,697 | 26,207 |
| Sweden | 40,118 | 25,810 | 25,773 | 25,836 | 26,167 | 25,669 | 25,843 |
| Poland | 38,886 | 30,819 | 30,902 | 30,946 | 31,026 | 31,242 | 30,871 |
| Belgium | 34,518 | 21,799 | 21,578 | 21,508 | 21,877 | 21,755 | 22,078 |
| Iran | 31,998 | 28,900 | 28,783 | 28,724 | 28,515 | 29,869 | 29,160 |
| Scotland | 25,553 | 14,691 | 14,740 | 14,767 | 15,356 | 14,500 | 14,700 |
| Israel | 24,156 | 17,667 | 17,682 | 17,658 | 17,868 | 18,030 | 18,038 |
| Denmark | 23,493 | 14,752 | 14,795 | 14,808 | 15,073 | 14,604 | 14,678 |

Table A2. MNCS of 25 countries according to seven counting methods.

| | FUC | AULFRC | ADLFRC | ORLFRC | COLFRC | FAC | CAC |
|---|---|---|---|---|---|---|---|
| United States | 1.34 | 1.31 | 1.30 | 1.30 | 1.29 | 1.32 | 1.32 |
| China | 0.95 | 0.88 | 0.88 | 0.88 | 0.88 | 0.89 | 0.88 |
| Germany | 1.25 | 1.10 | 1.09 | 1.09 | 1.09 | 1.10 | 1.10 |
| England | 1.42 | 1.28 | 1.27 | 1.27 | 1.27 | 1.30 | 1.30 |
| Japan | 0.87 | 0.78 | 0.79 | 0.79 | 0.79 | 0.78 | 0.78 |
| France | 1.19 | 1.02 | 1.02 | 1.02 | 1.02 | 1.03 | 1.02 |
| Canada | 1.28 | 1.12 | 1.12 | 1.12 | 1.14 | 1.12 | 1.13 |
| Italy | 1.18 | 0.99 | 1.00 | 1.00 | 1.00 | 0.99 | 0.99 |
| Spain | 1.16 | 0.99 | 0.99 | 1.00 | 1.00 | 0.99 | 0.99 |
| India | 0.74 | 0.67 | 0.67 | 0.68 | 0.68 | 0.68 | 0.67 |
| Australia | 1.30 | 1.15 | 1.16 | 1.16 | 1.16 | 1.16 | 1.16 |
| South Korea | 0.88 | 0.78 | 0.79 | 0.79 | 0.79 | 0.78 | 0.77 |
| Brazil | 0.72 | 0.62 | 0.62 | 0.62 | 0.62 | 0.62 | 0.61 |
| Netherlands | 1.54 | 1.36 | 1.36 | 1.36 | 1.37 | 1.38 | 1.37 |
| Russia | 0.50 | 0.34 | 0.34 | 0.35 | 0.35 | 0.33 | 0.31 |
| Taiwan | 0.92 | 0.85 | 0.86 | 0.86 | 0.86 | 0.85 | 0.85 |
| Turkey | 0.69 | 0.63 | 0.63 | 0.63 | 0.63 | 0.63 | 0.63 |
| Switzerland | 1.57 | 1.35 | 1.34 | 1.34 | 1.35 | 1.35 | 1.35 |
| Sweden | 1.38 | 1.13 | 1.14 | 1.14 | 1.16 | 1.13 | 1.12 |
| Poland | 0.72 | 0.54 | 0.55 | 0.55 | 0.56 | 0.54 | 0.53 |
| Belgium | 1.43 | 1.20 | 1.19 | 1.19 | 1.21 | 1.21 | 1.21 |
| Iran | 0.80 | 0.77 | 0.77 | 0.77 | 0.77 | 0.78 | 0.77 |
| Scotland | 1.50 | 1.27 | 1.26 | 1.27 | 1.29 | 1.29 | 1.28 |
| Israel | 1.16 | 0.96 | 0.97 | 0.97 | 0.99 | 0.96 | 0.96 |
| Denmark | 1.52 | 1.30 | 1.30 | 1.30 | 1.32 | 1.29 | 1.29 |

Table A3. $PP_{top\ 10\%}$ of 25 countries according to seven counting methods.

|  | FUC | AULFRC | ADLFRC | ORLFRC | COLFRC | FAC | CAC |
|---|---|---|---|---|---|---|---|
| United States | 15.0% | 14.6% | 14.5% | 14.4% | 14.2% | 14.6% | 14.7% |
| China | 9.3% | 8.4% | 8.5% | 8.5% | 8.5% | 8.5% | 8.5% |
| Germany | 13.7% | 11.6% | 11.4% | 11.5% | 11.5% | 11.7% | 11.6% |
| England | 15.9% | 14.0% | 13.9% | 13.9% | 13.9% | 14.2% | 14.3% |
| Japan | 7.6% | 6.3% | 6.4% | 6.4% | 6.5% | 6.2% | 6.2% |
| France | 12.7% | 10.3% | 10.3% | 10.3% | 10.2% | 10.4% | 10.3% |
| Canada | 13.6% | 11.5% | 11.5% | 11.5% | 11.7% | 11.5% | 11.6% |
| Italy | 12.1% | 9.4% | 9.5% | 9.5% | 9.5% | 9.4% | 9.3% |
| Spain | 11.8% | 9.6% | 9.6% | 9.6% | 9.7% | 9.6% | 9.6% |
| India | 6.1% | 5.3% | 5.3% | 5.3% | 5.3% | 5.3% | 5.2% |
| Australia | 14.0% | 12.0% | 12.0% | 12.0% | 12.1% | 12.1% | 12.0% |
| South Korea | 7.7% | 6.5% | 6.6% | 6.6% | 6.7% | 6.5% | 6.4% |
| Brazil | 5.5% | 3.9% | 4.0% | 4.0% | 4.1% | 3.9% | 3.8% |
| Netherlands | 17.6% | 15.1% | 15.0% | 15.0% | 15.1% | 15.4% | 15.4% |
| Russia | 3.7% | 1.7% | 1.9% | 1.9% | 2.0% | 1.7% | 1.5% |
| Taiwan | 8.3% | 7.4% | 7.5% | 7.5% | 7.5% | 7.4% | 7.4% |
| Turkey | 5.5% | 4.7% | 4.7% | 4.7% | 4.7% | 4.7% | 4.6% |
| Switzerland | 18.3% | 15.7% | 15.5% | 15.5% | 15.7% | 16.0% | 15.9% |
| Sweden | 14.7% | 11.3% | 11.4% | 11.4% | 11.6% | 11.3% | 11.1% |
| Poland | 5.7% | 3.4% | 3.5% | 3.5% | 3.6% | 3.3% | 3.1% |
| Belgium | 15.8% | 12.8% | 12.7% | 12.7% | 12.9% | 13.0% | 13.0% |
| Iran | 7.4% | 6.9% | 6.9% | 6.9% | 6.9% | 7.0% | 6.9% |
| Scotland | 16.8% | 13.7% | 13.6% | 13.6% | 14.1% | 14.0% | 13.9% |
| Israel | 11.8% | 9.1% | 9.1% | 9.1% | 9.4% | 9.1% | 9.0% |
| Denmark | 17.2% | 14.2% | 14.2% | 14.2% | 14.4% | 14.2% | 14.1% |